\documentclass[twocolumn,aps,showpacs,prb,tightenlines,amsmath,amssymb,superscriptaddress]{revtex4}
\usepackage{graphicx}
\usepackage{amssymb}
\usepackage{dcolumn} 
\usepackage{amsmath}
\usepackage{bm}
\usepackage{colordvi}
\newcommand{\bgreek}[1]{\mbox{\boldmath$#1$\unboldmath}}
\makeatletter

\newcommand{\Rmnum}[1]{\expandafter\@slowromancap\romannumeral #1@}
\makeatother

\begin{document}

\title{Voltage controlled spin precession in InAs quantum wells}
\author{B. Y. Sun}
\affiliation{Hefei National Laboratory for Physical Sciences at
  Microscale,
University of Science and Technology of China, Hefei,
  Anhui, 230026, China}
\affiliation{Department of Physics,
University of Science and Technology of China, Hefei,
  Anhui, 230026, China}
\author{P. Zhang}
\affiliation{Department of Physics,
University of Science and Technology of China, Hefei,
  Anhui, 230026, China}
\author{M. W. Wu}
\thanks{Author to  whom correspondence should be addressed}
\email{mwwu@ustc.edu.cn.}
\affiliation{Hefei National Laboratory for Physical Sciences at
  Microscale,
University of Science and Technology of China, Hefei,
  Anhui, 230026, China}
\affiliation{Department of Physics,
University of Science and Technology of China, Hefei,
  Anhui, 230026, China}
\date{\today}

\begin{abstract}
 In this work we investigate spin diffusion in InAs
   quantum wells with the Rashba spin-orbit coupling modulated by a
   gate voltage. The gate voltage dependence of the
   spin diffusion under different temperatures is studied with all the
   scattering explicitly included. Our result partially supports the claim of
   the realization of the Datta-Das spin-injected field effect
   transistor by Koo {\it et al.} [Science {\bf 325}, 1515 (2009)]. 
   We also show that the scattering plays an important role in spin
   diffusion in such a system.
\end{abstract}


\maketitle

In the past decades, a great deal of effort has been made for the
realization of the spintronic devices.\cite{Awschalom,Zutic,Dyakonov,wuReview} 
The spin-injected field effect transistor (SIFET), proposed by
Datta and Das in 1990,\cite{DattaDas} is one of
the most intriguing devices\cite{Pala,Gelabert,Sahoo} but posts some 
challenges to experiments (e.g., the spin-polarized injection and
detection). Very recently Koo {\it et al.}\cite{Koo} reported that, in InAs
quantum wells with nonlocal spin valve configuration, the nonlocal voltage was
observed to oscillate with the variation of gate voltage at the low temperature when
the two ferromagnetic electrodes (spin injector and detector) are magnetized
along the spin diffusion direction. They claimed that they have realized the
SIFET because the oscillation can be fitted by a theoretical equation
describing the SIFET. 
Nevertheless, as pointed out by Bandyopadhyay,\cite{Bandyopadhyay}
the theoretical equation adopted by Koo {\it et
  al.} only applies to the one-dimensional system instead of the two-dimensional
one. Therefore the 
agreement between this equation and the experimental data\cite{Koo}
makes little meaning and doubt is cast on the conclusion presented by Koo
{\it et al.}. Later, Zainuddin {\it et al.}\cite{Zainuddin}
extended the one-dimensional theory to the two-dimensional
case with an equation similar to the one obtained from the one-dimensional
theory. However, as further reported by Agnihotri and
Bandyopadhyay,\cite{Agnihotri} the experimental data actually do not
match the equation for the two-dimension SIFET. Therefore, whether the device
proposed by Koo {\sl et al.} realizes the SIFET is still under
debate. It is noted that all the theoretical works mentioned above
were performed without any scattering. However, the scattering
  exists in reality and can be very important for spin
  diffusion.\cite{wuReview,JLCheng,Zhang}

In fact, a thorough understanding of spin diffusion in the two-dimensional SIFET
with the scattering explicitly included can be obtained based on the kinetic spin Bloch
equation (KSBE) approach,\cite{wuReview} which has been successfully applied to
study the spin diffusion/transport in various two-dimensional systems (e.g.,
GaAs quantum wells\cite{weng,LJiang,JLCheng,ani} and Si/SiGe quantum
wells\cite{Zhang}). In the framework of this approach,  spins of 
electrons with wave-vector ${\bf k}$ precess in spatial domain with frequency
\begin{equation}
{\bgreek\omega_{\bf k}}=m^\ast({\bf \Omega_k}+g\mu_B{\bf B})/k_x
\label{frequency}
\end{equation}
during the spin diffusion.\cite{weng,JLCheng} Here, the spin diffusion
direction is set to be the ${\bf \hat x}$-axis, $m^\ast$ is the effective
electron mass, ${\bf \Omega_k}$ is the D'yakonov-Perel' (DP)\cite{DP}
spin-orbit coupling term and ${\bf B}$ is the external magnetic field. 
In InAs quantum wells, the Rashba spin-orbit coupling\cite{Rashba}
dominates and thus ${\bgreek \Omega}_{\bf k}=2\alpha(-k_y,k_x,0)$ with $\alpha$
being the Rashba coefficient modulated by the gate voltage.
Moreover, the small external magnetic field used to magnetize
the electrodes can be neglected when compared to the Rashba spin-orbit
coupling.\cite{Koo} Therefore, the spatial spin precession frequency
\begin{equation}
{\bgreek\omega_{\bf k}}=2\alpha m^\ast(-\tan\theta_{\bf k},1,0)
\label{omegak}
\end{equation}
depends on the polar angle $\theta_{\bf k}$ of the momentum. This
  ${\bf k}$ dependence of the precession frequency leads to the inhomogeneous
broadening.\cite{weng,wu1} The inhomogeneous broadening itself causes reversible
spin relaxation during spin diffusion.\cite{wuReview} One notices that the
one-dimension model adopted by Koo {\it et al.}\cite{Koo} actually excludes the
inhomogeneous broadening by neglecting the transverse component of the momentum
(i.e., $k_y$) and therefore is inappropriate. The scattering also plays an
 important role in spin diffusion which makes the relaxation irreversible and
affects the spin diffusion length or even the precession
frequency.\cite{JLCheng,Zhang} Furthermore, the temperature ($T$) dependence of
the spin diffusion length should be mainly from the temperature dependence 
of the scattering in this case as
the inhomogeneous broadening is insensitive on $T$
($\omega_{\bf k}$ does not depend on the magnitude of ${\bf k}$). In this
paper, we numerically solve the KSBEs under the DP
mechanism and obtain the gate-voltage dependence of spin polarization
at the detection point. Our result is in good agreement with the
experiment of Koo {\it et al.} at low and sufficiently high
temperatures and hence  is in favor of their claim of the
realization of the SIFET. Moreover, the role played by the scattering
in spin diffusion is also investigated and revealed to be important.

We start our investigation from InAs quantum wells as presented in
Ref.\,\onlinecite{Koo}. The depth $V_0$ and width $a$ of the square well are set
to be 430~meV and 2~nm, respectively. The initial spatially uniform electron density
$N_e$ is 2.5$\times$10$^{12}$~cm$^{-2}$ and the effective electron mass $m^\ast= 0.05m_0$
where $m_0$ is the free electron mass. The ${\bf\hat x}$-axis
polarized spins (the polarization $P_0$ is set to be 0.02) are
injected at the left boundary $x=0$ and diffuse 
along the ${\bf\hat x}$-axis. Due to the narrow well width, moderate
electron density and small polarization, only the lowest
subband is relevant in our investigation. The Rashba spin-orbit coupling
coefficient $\alpha$ is taken from Koo {\it et al.}.\cite{Koo} The impurity 
density $N_i$ is estimated to be 0.039$N_e$ according to the mobility
$\mu=50000$~cm$^2$ V$^{-1}$ s$^{-1}$ reported by
Koo {\it et al.}. It is noted that these
parameters (e.g., $N_i$ and 
  $N_e$, etc.) are obtained from the low temperature case of the
  experiment of Koo {\it et al.}. Although these parameters may vary with the
  temperature,\cite{Kwon} we still apply them to the high temperature
  investigations due to the lack of necessary experimental information. The
  other parameters can be found in Ref.\,\onlinecite{InAs}. The KSBEs
  read 
\begin{eqnarray}
  \frac{\partial\rho_{\bf k}(x,t)}{\partial
    t}&&=-e\frac{\partial \Psi(x,t)}{\partial x}\frac{\partial \rho_{\bf
      k}(x,t)}{\partial k_x}-\frac{k_x}{m^\ast}\frac{\partial \rho_{\bf
      k}(x,t)}{\partial x}  \nonumber \\ 
  &&\mbox{}-i[{\bgreek \Omega}_{\bf k}\cdot\frac{{\bgreek
      \sigma}}{2},\rho_{\bf k}(x,t)] 
  +\frac{\partial\rho_{\bf k}(x,t)}{\partial t}\Big|_{\mbox{scat}}.
\end{eqnarray}
Here, $\rho_{\bf k}(x,t)$ are the single-particle density matrices of electrons
with the in-plane wave-vector ${\bf k}$ at position $x$ and time $t$. $\Psi(x,t)$ is the electric potential
satisfying the Poisson equation $\nabla^2_x\Psi(x,t)=e[n(x,t)-N_0]/(a
\kappa_0 \varepsilon_0)$ with $n(x,t)=\sum_{\bf k}$Tr$[\rho_{\bf k}(x,t)]$ standing for the electron density at
position $x$ and time $t$, $N_0$ the background positive charge
density, and $\kappa_0$ the relative static dielectric constant. $-i[{\bgreek \Omega}_{\bf
    k}\cdot{\bgreek \sigma}/2,\rho_{\bf k}(x,t)]$ is the coherent term describing
the spin precession. $\frac{\partial \rho_{\bf k}(x,t)}{\partial
  t}\big|_{\mbox{scat}}$ is the scattering term with the
electron-impurity, electron-acoustic/longitudinal optical
phonon, and electron-electron scatterings 
included. The details of the scattering term can be found
in Refs.\,\onlinecite{wuReview}, \onlinecite{weng2} and
  \onlinecite{Zhou}. It is noted that no
  fitting parameter is needed in our calculation.

To solve the KSBEs,  the initial conditions are set as
\begin{eqnarray}
&&\rho_{\bf k}(0,0)=(F^0_{{\bf k}\uparrow}+F^0_{{\bf
      k}\downarrow})/2+(F^0_{{\bf k}\uparrow}-F^0_{{\bf
      k}\downarrow})\sigma_x/2,\\
&&\rho_{\bf k}(x>0,0)=(F^L_{{\bf k}\uparrow}+F^L_{{\bf
      k}\downarrow})/2,
\end{eqnarray}
and the boundary conditions are given as\cite{JLCheng}
\begin{eqnarray}
&&\hspace{-0.15cm}\rho_{\bf k}(0,t)|_{k_x>0}=(F_{{\bf k}\uparrow}^0+F_{{\bf
    k}\downarrow}^0)/2+(F_{{\bf k}\uparrow}^0-F_{{\bf k}\downarrow}^0){\sigma_x}/2,\\
&&\hspace{-0.15cm}\rho_{\bf k}(L,t)|_{k_x<0}=(F_{{\bf k}\uparrow}^L+F_{{\bf
    k}\downarrow}^L)/2,\\
&&\hspace{-0.15cm}\Psi(0,t)=\Psi(L,t)=0.
\end{eqnarray}
Here, $x=L$ stands for the right boundary with $L$ much
longer than the spin diffusion length. $F^{0,L}_{{\bf
    k}\uparrow}$ ($F^{0,L}_{{\bf  k}\downarrow}$) stand for the
Fermi distributions of electrons with spin parallel (antiparallel) to the
${\bf\hat x}$-axis determined by the temperature and the
initial polarization at the two boundaries. The numerical scheme for solving the
KSBEs can be found in detail in
Ref.\,\onlinecite{JLCheng}. With the single-particle density matrices obtained
by solving the KSBEs, the spin polarization at the point $x$ at the steady state
can be obtained as 
\begin{eqnarray}\nonumber
  P(x,+\infty)&=&\sum_{\bf k}\mbox{Tr}[\rho_{\bf
    k}(x,+\infty)\sigma_x]/n(x,+\infty)\\
  &\equiv&\sum_{\bf k}P_{\bf k}(x,+\infty).
  \label{Polar}
\end{eqnarray}
Since the nonlocal voltage measured in the experiment is proportional to the
spin polarization at the detection point,\cite{Salis,Huang} we fit the
experimental data measured at $x_0$ with $P(x_0,+\infty)$.

\begin{figure}[htb]
\centering
\includegraphics[width=6cm]{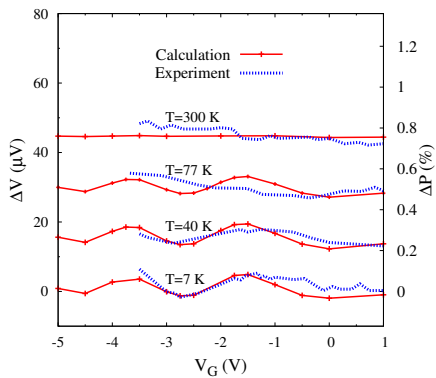}
\caption{(Color online) Gate voltage dependence of the spin polarization obtained
  from the KSBEs (solid  curves with the scale on the right hand side of the frame)
  and that of the nonlocal voltage measured in the experiment\cite{Koo}  
  (dashed curves) at the detection point $x_0=1.25$~$\mu$m under different
  temperatures. The plots are shifted for clarity as Koo {\em et al.}.\cite{Koo}}
\label{figszw1}
\end{figure}

In Fig.~\ref{figszw1}, we plot the gate voltage $V_G$ dependence of the spin  
polarization at the detection point $x_0=1.25$~$\mu$m by the solid
curves and that of the experimentally
measured nonlocal voltage by the dashed curves under different
temperatures. It is noted that the Rashba
  spin-orbit coupling coefficient used in our calculation with the gate voltage
  from $-5$~V to $-3$~V is obtained by linearly extending the $\alpha$-$V_G$ curve
  presented by Koo {\it et al.}.\cite{Koo} From 
Fig.~\ref{figszw1}, one finds that our result is in good agreement 
with the experiment at 7~K and 40~K. Moreover, more than one
  period of oscillation are found in our result, suggesting
that Koo {\it et al.} may also observe more periods if they enlarge
the scope of measurement. When the temperature is higher, more
  impurities will be ionized and both the electron and impurity
  densities will increase.\cite{Kwon} This may explain the discrepancy between our
  theoretical result and the experimental data at $T=77$~K. When
  $T=300$~K and hence the scattering is strong, both our calculation and the
  experiment by Koo {\it et al.} show that the oscillation of the spin
  polarization/nonlocal voltage disappears. In fact, due to the
  suppression on spin diffusion (mainly caused by the strengthened
  scattering, as revealed in the following), the spin diffusion length
  becomes much shorter than the spacing between the injector and detector,
  and therefore no spin polarized signal can be observed at the detection point.

To further investigate the influence of scattering on the spin
diffusion, we first consider the much simplified case without the 
scattering and electric field, the single-particle density matrix for
any ${\bf k}$ in the steady state can be obtained easily from the KSBEs 
as\cite{JLCheng}
\begin{eqnarray}
  \rho_{\bf k}(x,+\infty)=
  \begin{cases}
    e^{\frac{-i{\bgreek \omega_{\bf k}}\cdot{\bgreek \sigma}}{2}x}\rho_{\bf
      k}(0,0)e^{\frac{i{\bgreek \omega_{\bf k}}\cdot{\bgreek\sigma}}{2}x},& \hspace{-0.1 cm}k_x>0\\
    \rho_{\bf k}(L,0),& \hspace{-0.1 cm} k_x<0
  \end{cases}
  ,
\label{noscat}
\end{eqnarray}
where $\bgreek \omega_{\bf k}$ is given in Eq.~(\ref{omegak}).
Then at the detection point,
\begin{eqnarray}
\hspace{-0.2 cm}P_{\bf k}(x_0,+\infty)=
 \begin{cases}
   B_{\bf k}[s^2+(1-s^2)\cos(\frac{\theta_{x_0}}{\sqrt{1-s^2}})],&
   \hspace{-0.3 cm}k_x>0\\
   0,& \hspace{-0.3 cm}k_x<0
  \end{cases}
\end{eqnarray}
 with $s=k_y/k=\sin\theta_{\bf k}$, $\theta_{x_0}=2m^\ast\alpha
 x_0$ and $B_{\bf k}=(F_{{\bf k}\uparrow}^0-F_{{\bf
     k}\downarrow}^0)/N_e$. This solution with $k_x>0$ has the same form as the
 result from Zainuddin {\it et al.} [Eq.~(5a) in
 Ref.\,\onlinecite{Zainuddin}]. Our result clearly indicates that the
 contribution to the total spin-polarized signal mainly comes from the
 $k_x$-positive states around the Fermi circle. Instead of summing
 $P_{\bf k}(x_0,+\infty)$ over the $k_x$-positive Fermi circle line as
 done by Zainuddin {\it et al.},\cite{Zainuddin} we take into account
 all the $k_x$-positive states and obtain
\begin{eqnarray}
  P(x_0,+\infty)\propto\int_{-\frac{\pi}{2}}^{\frac{\pi}{2}}d\theta_{\bf k}(1-2\sin^2\frac{m^\ast\alpha x_0}{\cos\theta_{\bf k}}\cos^2\theta_{\bf
      k}).
  \label{Ptheory}
\end{eqnarray}
It is noted that the integration over $\theta_{\bf k}$ in Eq.~(\ref{Ptheory}) stands
for the interference among different ${\bf k}$ states. However, one finds that
  this equation can not fit the experimental data very well as the
  situation faced by Agnihotri and Bandyopadhyay\cite{Agnihotri}
   until the scattering is included as presented previously.

\begin{figure}[htb]
\centering
\includegraphics[width=6cm]{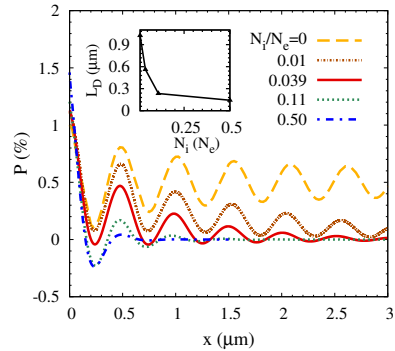}
\caption{(Color online) $x$ dependence of $P$ with different
  impurity densities. The impurity density dependence of the spin
  diffusion length is also plotted in the inset. The case of
$N_i=0.039N_e$ corresponds to the experimental situation. The
temperature is 7~K and the gate voltage is 0.} 
\label{figszw2}
\end{figure}

We then solve the KSBEs numerically by varying the impurity density
artificially. Without losing generality, we take the temperature to be
7~K and the gate voltage to be 0. Under these conditions, the $x$
dependence of the spin polarization with different impurity densities
are plotted in Fig.~\ref{figszw2}. It is noted that the curve
  with $N_i=0$ in this figure is obtained directly from
  Eq.~(\ref{Ptheory}).
The impurity density dependence of the spin diffusion length is plotted in
  the inset of Fig.~\ref{figszw2}. From this inset, one 
finds that the spin diffusion length decreases sensitively with the increase
in the impurity density. It is noted that even for the
case of $N_i=0.5N_e$, the 
system is still in the weak scattering limit as $\omega_L\tau_p=1.19>1$, where
$\omega_L=2\alpha k_F$ is the spin precession frequency due to the
spin-orbit coupling and $\tau_p$ is the momentum relaxation time. The decrease
in the spin-diffusion length with the increase in the impurity density can be
understood alternatively by means of the quasi-independent electron 
model,\cite{Ziese,Flatte,Zutic2,Martin} where the spin diffusion 
length is characterized by $\sqrt{D_s\tau_s}$ with $\tau_s$ standing for the
spin relaxation time and $D_s$ representing the spin diffusion constant. $D_s$
decreases with the increasing scattering strength.\cite{wuReview} 
$\tau_s$ has the same tendency as $D_s$ as long as  electrons are in the weak
scattering limit.\cite{wuReview} Therefore, the spin-diffusion length
decreases with the increase in scattering
  strength. This explains the disappearance of the oscillation
  of the spin-polarized signal at $T=300$~K in Fig.~\ref{figszw1} since the
  electron-phonon scattering is strengthened there.

In summary, we have investigated the spin diffusion in $n$-type InAs quantum
wells with the scattering explicitly included under the DP
mechanism. The consistency between our theoretical result and the
experimental data partially supports the
 claim by  Koo {\it et al.}\cite{Koo} that 
a SIFET has been demonstrated.
The essential role played by the scattering is
also revealed. It is shown that the spin diffusion length decreases
with the increase in the impurity density in the weak scattering limit.

This work was supported by the Natural Science Foundation of China
under Grant No.\ 10725417.


\begin{thebibliography}{0}
\bibitem{Awschalom} {\it Semiconductor Spintronics and Quantum Computation},
  edited by D. D. Awschalom, D. Loss, and N. Samarth (Sprinter, Berlin, 2002);
  and references therein.
\bibitem{Zutic} I. $\check{\mbox{Z}}$uti$\acute{\mbox{c}}$, J. Fabian, and
  S. Das Sarma, Rev. Mod. Phys. {\bf 76}, 323 (2004); and references therein.
\bibitem{Dyakonov} {\it Spin Physics in Semiconductors}, ed. by M. I. D'yakonov
  (Springer, Berlin, 2008); and references therein.
\bibitem{wuReview} M. W. Wu, J. H. Jiang, and M. Q. Weng, Phys. Rep. {\bf 493},
  61 (2010); and references therein.
\bibitem{DattaDas} S. Datta and B. Das, Appl. Phys. Lett. {\bf 56}, 665 (1990).

\bibitem{Pala} M. G. Pala, M. Governale, J. K\"{o}nig, and
  U. Z\"{u}licke, Europhys. Lett. {\bf 65}, 850 (2004).

\bibitem{Sahoo} S. Sahoo, T. Kontos, J. Furer, C. Hoffmann,
  M. Gr\"{a}ber, and C. Sch\"{o}nenberger, Nat. Phys. {\bf 1}, 99 (2005).


\bibitem{Gelabert} M. M. Gelabert, L. Serra, D. S\'{a}nchez, and
  R. L\'{o}pez, Phys. Rev. B {\bf 81}, 165317 (2010).

\bibitem{Koo} H. C. Koo, J. H. Kwon, J. Eom, J. Chang, S. H. Han, and
  M. Johnson, Science {\bf 325}, 1515 (2009).
\bibitem{Bandyopadhyay} S. Bandyopadhyay, arXiv:0911.0210v1
\bibitem{Zainuddin} A. N. M. Zainuddin, S. Hong, L. Siddiqui, and S. Datta,
  arXiv:1001.1523v2
\bibitem{Agnihotri} P. Agnihotri and S. Bandyopadhyay, Physica E {\bf 42}, 1736
  (2010).
\bibitem{JLCheng} J. L. Cheng and M. W. Wu, J. Appl. Phys. {\bf 101}, 073702
  (2007).
\bibitem{Zhang} P. Zhang and M. W. Wu, Phys. Rev. B {\bf 79}, 075303 (2009).
\bibitem{weng} M. Q. Weng and M. W. Wu, Phys. Rev. B {\bf 66}, 235109 (2002).

\bibitem{LJiang} L. Jiang, M. Q. Weng, M. W. Wu, and J. L. Cheng,
  J. Appl. Phys. {\bf 98}, 113702 (2005).

\bibitem{ani}J. L. Cheng, M. W. Wu, and I. C. da Cunha Lima, Phys. Rev. B 
{\bf 75}, 205328 (2007). 

\bibitem{DP} M. I. D'yakonov and V. I. Perel', Zh. Eksp. Teor. Fiz. {\bf 60},
  1954 (1971) [Sov. Phys. JETP {\bf 33}, 1053 (1971)].
\bibitem{Rashba} Y. A. Bychkov and E. I. Rashba, J. Phys. C {\bf 17}, 6039
  (1984); Pis'ma Zh. \'Eksp. Teor. Fiz. {\bf 39}, 66 (1984) [JETP Lett. {\bf
    39}, 78 (1984)].
\bibitem{wu1} M. W. Wu and C. Z. Ning, Eur. Phys. J. B {\bf 18}, 373 (2000);
  M. W. Wu, J. Phys. Soc. Jpn. {\bf 70}, 2195 (2001).
\bibitem{Kwon} J. H. Kwon, H. C. Koo, J. Chang, and S. H. Han,
  Appl. Phys. Lett. {\bf 90}, 112505 (2007).
\bibitem{InAs} J. H. Jiang and M. W. Wu, Phys. Rev. B {\bf 79}, 125206
  (2009).

\bibitem{weng2} M. Q. Weng, M. W. Wu, and L. Jiang, Phys. Rev. B {\bf 69},
  245320 (2004).
\bibitem{Zhou} J. Zhou, J. L. Cheng, and M. W. Wu, Phys. Rev. B {\bf 75}, 045305
  (2007).
\bibitem{Salis} G. Salis, A. Fuhrer, R. R. Schlittler, L. Gross, and
  S. F. Alvarado, Phys. Rev. B {\bf 81}, 205323 (2010).
\bibitem{Huang} B. Huang, D. J. Monsma, and I. Appelbaum, 
Phys. Rev. Lett. {\bf 99}, 177209 (2007).

\bibitem{Ziese} {\it Spin Electronics}, edited by M. Ziese and M. J. Thornton (Springer,
  Berlin, 2001).
\bibitem{Flatte} M. E. Flatt\'e and J. M. Byers, Phys. Rev. Lett. {\bf 84}, 4220
  (2000).
\bibitem{Zutic2} I. \v{Z}uti\'c, J. Fabian, and S. Das Sarma, Phys. Rev. B {\bf 64},
  121201 (2001).
\bibitem{Martin} I. Martin, Phys. Rev. B {\bf 67}, 014421 (2003).
\end{thebibliography}
\end{document}